# Manipulating terahertz phonon-polariton in the ultrastrong coupling regime with bound states in the continuum

*Jiaxing Yang[1#], Liyu Zhang[1#], Kai Wang[1,2*], Chen Zhang[1], Aoyu Fan[3], Zijian He[1], Zhidi Li[1], Xiaobo Han[4], Furi Ling[3*], Peixiang Lu[1,4*]*

[1]Wuhan National Laboratory for Optoelectronics and School of Physics, Huazhong University of Science and Technology, Wuhan 430074, China

[2]School of Electronic and Information Engineering, Hubei University of Science and Technology, Xianning 437100, China

[3]School of Optics and Electronic Information, Huazhong University of Science and Technology Wuhan, Hubei 430074, P.R. China

[4]Hubei Key Laboratory of Optical Information and Pattern Recognition, Wuhan Institute of Technology, Wuhan 430205, China

*Corresponding authors:

#These authors contributed equally to this work.

kale_wong@hust.edu.cn (KW), lingfuri@mail.hust.edu.cn (FRL), lupeixiang@hust.edu.cn (PXL)

**Abstract**

The strong coupling between photons and phonons in polar materials gives rise to phonon-polaritons that encapsulate a wealth of physical information, offering crucial tools for the ultrafast terahertz sources and the topological engineering of terahertz light. However, it is still quite challenging to form and manipulate the terahertz phonon-polaritons under the ultrastrong coupling regime till now. In this work, we demonstrate the ultrastrong coupling between the phonon (at 0.95 THz) in a $MAPbI_3$ film and the metallic bound states in the continuum (BICs) in Au metasurfaces. The Rabi splitting can be continuously tuned from 28% to 48.4% of the phonon frequency by adjusting the parameters (size, shape and period) of Au metasurfaces, reaching the ultrastrong coupling regime. By introducing wavelet transform, the mode evolution information of the terahertz phonon-polariton is successfully extracted. It indicates that the phonon radiation intensity of the $MAPbI_3$ film is enhanced as the coupling strength is increased. This work not only establishes a new platform for terahertz devices but also opens new avenues for exploring the intricate dynamics of terahertz phonon-polaritons.

**Introduction**

Phonon-polariton is the quasi-particle generated from the interaction between light and phonons in materials, which is widely distributed from mid-infrared to terahertz (THz) ranges [1–6]. It carries crucial physical information and exhibits potential applications, such as vacuum Bloch–Siegert shift[7], topological control [8], tunable laser sources [9] and thermal emitters [2]. Notably, terahertz phonon-polariton is formed through the coupling of photons with THz phonons, which is a vital research focus within the realm of polaritons recently. Since the THz phonons widely exist in many prominent materials such as semiconductor quantum wells[10], perovskites [11], graphene [5] and DNA molecules [12], it can provide deep insights into these materials and crucial guidance for the design of terahertz sensors [13] and detectors [14,15]. Therefore, it is essential to develop a hybrid system for THz phonon-polaritons with an enhanced coupling strength. The coupling strength in such systems is quantified by $\eta = \Omega_R/2\omega_{Ph}$, where $\Omega_R$ represents the vacuum Rabi frequency and $\omega_{Ph}$ the photon frequency. Especially for the hybrid system operating in the ultrastrong coupling (USC) regime (η>0.1) [16–19], it transcends conventional physical limits and exhibits unique phenomena [20–22]. The USC regime has garnered significant attention due to its demonstrated potential for modifying material properties, include modified electronic transport [23,24], cavity chemistry [25,26], or vacuum-field-induced super-conductivity [27,28].

Most previous studies on terahertz phonon-polaritons are based on Fabry-Perot (F-P) cavities [29–33] or plasmonic nanocavities [34–38]. Specifically, F-P cavities have a large mode volume with uniform local-field distributions, making it convenient for coupling with the thin films of organic molecules [32] or crystalline semiconductor [39]. However, the local-field enhancement in F-P cavities is moderate, leading to the limitation in achieving ultrastrong coupling strength. Furthermore, it is still challenging for miniaturization and integration. In contrast, plasmonic nanocavities offer a stronger local-field in a much smaller mode volume in nanostructures [40], thereby enhancing the coupling strength of the system. However, the typical mode volume of plasmonic nanocavities is much smaller than crystalline particles, such as MAPbI$_3$ (~ 300 nm) [41], α-lactose (> 10 $\mu$m) [42], leading to a spatial mismatching between the cavity and the phonon modes. As mentioned, both cavities show the limitations for further enhancing the coupling strength of THz phonon-polaritons. Therefore, it is urgent to develop a new hybrid system for studying THz phonon-polaritons under the ultrastrong coupling regime.

In recent years, bound states in the continuum (BICs) have been actively studied in optical systems [43,44], accompanied by many discoveries and applications such as optical microcavities [45], lasers [46] and sensors [47,48]. BICs have been shown the excellent abilities for enhancing light-matter interactions [49,50], providing a strongly enhanced local-field, high quality factor ($Q$), ultrathin thickness, and broad resonance tunability via variation of their geometrical parameters [51]. In particular, BICs can reduce thermal losses while simultaneously increasing coupling strength in the system of metallic metasurfaces [52,53]. These capabilities make metallic metasurfaces based on BICs a promising platform for achieving and modulating ultrastrong coupling in the terahertz range. In addition, MAPbI$_3$ is a perovskite material with excellent optoelectronic properties, which shows potential applications in batteries [54], solar energy conversion [55], and light-emitting diodes (LEDs) [56]. MAPbI$_3$ exhibits strong phonon vibrations at 0.95 THz, and can be easily crystallized into a high-quality crystalline film through spin-coating and thermal annealing making them a highly significant terahertz material that has attracted considerable attention [57–60]. Therefore, MAPbI$_3$ is a suitable phonon material for THz phonon-polaritons.

Current investigations of USC in terahertz phonon-polariton systems predominantly rely on

Fourier-transform spectroscopy to characterize stationary spectral anti-crossing features. However, this approach obscures critical dynamical information, particularly the real-time mode evolution and phase-resolved intensity variations in phonon-polaritons. Such limitations obstruct the experimental exploration of dynamics process in USC systems. Wavelet analysis emerges as a powerful mathematical framework for resolving localized time-frequency characteristics of non-stationary signals. This technique captures joint temporal-spectral information using scalable, translatable wavelet basis functions, making it suitable for probing transient processes in USC systems. The generalized Morse wavelet transform is a parametrically optimized analytic wavelet, which has demonstrated exceptional performance in telecommunications and biomedical signal processing. This methodology can provide multidimensional insights into terahertz polariton systems, including coupling-strength modulation phase-discontinuity and other critical parameters which cannot be observed through conventional spectral integration techniques.

Here, we report the demonstration of the ultrastrong coupling between the phonon in $MAPbI_3$ film and the metallic BICs in Au metasurfaces. The unit cell of the Au metasurface consists of the coupled C-shaped Au split ring resonator (SRR) pairs, forming the BIC modes. The resonance linewidth of BICs is controlled via the asymmetry of the unit cell, matching with the damping rate of phonon vibration. By varying the size of unit cells, the BIC resonance frequency is continuously tuned to match with the $MAPbI_3$ phonon frequency at 0.95 THz, and a Rabi splitting up to 0.28 THz is obtained. Importantly, the Rabi splitting can further be tuned from 28% to 48.4% of the phonon frequency by precisely controlling the mode volume of BICs, reaching the ultrastrong coupling regime. By introducing wavelet transform, the mode evolution information of the terahertz phonon-polariton is successfully extracted. It indicates that the phonon radiation intensity of the $MAPbI_3$ film is enhanced as the coupling strength is increased. Our results open up new possibilities for the control of polaritons and reveal new information within the phonon-polaritons system.

**Results**

**1. Sample design and BICs theory**

Figure 1a illustrates the proposed hybrid system, which is composed of Au metasurface and MAPbI$_3$ film (see Method for details). Figure 1b shows the unit cell of the Au metasurface, it can be seen the asymmetric C-shaped Au SRR pairs, and the asymmetric degree can be controlled by the difference of arm lengths, $\Delta L = L_1 - L_2$. As the two electromagnetic modes in Au SRR pairs are coupled with each other, two new modes are generated. In special cases, one of these modes exhibits zero radiation loss, which is called BIC. The metallic BIC modes discussed in this paper can be elucidated using the Friedrich–Wintgen (F-W) BICs theory. In principle, BIC modes are analyzed by treating the metal as a perfect electric conductor (PEC), the simulated $Q$ factor exhibits an inverse square relationship with $\Delta L$ as shown by the orange curve in Figure 1c. For symmetric structures ($\Delta L = 0$) the resonance vanishes as the quasi-BIC turns into a true BIC, i.e., the $Q$ factor of the BIC modes approaches infinity, because the loss associated with the metal is considered to be zero in the PEC model. When the Drude model is used to depict the refraction index of Au, the simulated $Q$ factor of the BIC modes is shown by the blue curve in Figure 1c, which matches well with the PEC model when $\Delta L$ is large. However, as $\Delta L$ approaches zero, the $Q$ factor is limited to below 200, because of the inherent losses of Au in the Drude model. The red stars in Figure 1c represent the experimental results of the $Q$ factor which show a trend consistent with the Drude model (see Method for experiment setup). The relationship approximately follows an inverse square proportionality with respect to $\Delta L$. Due to factors such as uncertainties during the sample preparation, the measured $Q$ values are lower than the simulated results under the Drude model at low asymmetry (see Note 1, Supplementary Information).

MAPbI$_3$ is selected as the phonon material for the coupling system. In the classical model, the dielectric constant dispersion of phonon modes is typically described by Lorentz model. The relative permittivity can be expressed as [11]:

$$\varepsilon_\mathrm{r} = \varepsilon_\infty + \chi = \varepsilon_\infty - \sum_m \frac{S_{0m} * \omega_{0m}^2}{\omega^2 - \omega_{0m}^2 + i\omega\gamma_m} \qquad (1)$$

where $\chi$ is the polarization susceptibility, $\varepsilon_\infty$ is the permittivity at infinite frequency, $\omega_0$ and $\gamma$ represent the phonon frequency and damping rate. The refractive index is given by $n = \sqrt{\varepsilon_\mathrm{r}}$ (see Note 2, Supplementary Information).

## 2. Strong coupling in Au-MAPbI$_3$ hybrid metasurfaces

As shown in Figure 2a, the BIC resonance peak is continuously tuned from 0.68 to 1.53 THz through the phonon frequency at 0.95 THz by adjusting the scaling factor $S$ of the unit cell. $S$ parameter is defined as the proportional magnified ratio in size in comparison with the standard size of the unit cell in Figure 1b. Noting that the asymmetric parameter is kept as the constant of $\Delta L = 34$ μm, when the resonance linewidth of the BIC mode is matched with the damping rate of the phonon. At this parameter of asymmetry, the Rabi splitting due to the coupling between the BICs and phonons is most pronounced (see Note 3, Supplementary Information). Figure 2b shows the position of the BIC resonance peak before (red dashed line) and after (red solid line) MAPbI$_3$ film coating. It shows a significant redshift after coating, due to the influence of the vacuum dielectric constant of MAPbI$_3$. At this stage, the MAPbI$_3$ film has not undergone annealing and is in an intermediate phase, which exhibits no phonon resonance. Figure 2c shows the transmission amplitude of the crystallized MAPbI$_3$ film on a quartz substrate after thermal annealing. The quality of crystallization of perovskite significantly affects the coupling strength. The MAPbI$_3$ thin film has a thickness of approximately 200 nanometers and exhibits good crystallization quality, with a particle size significantly smaller than the minimum size of the metasurface unit cells. This allows for optimal coupling between the BICs and the phonons of MAPbI$_3$ (Figure S18a, Supplementary Information).

After crystallization of the perovskite films on the Au metasurfaces, as shown in Figure 2d, the transmission amplitude of the system exhibits a pronounced anti-crossing feature. To explain the changes in modes before and after coupling between BICs and phonons, the variations in the energy levels of the system and the interactions between these energy levels were analyzed. From the Hamiltonian of the system, the energy levels and interactions that govern the dynamics of the coupled modes within the hybrid metasurfaces can be derived. The full Hamiltonian can be expressed in the following form[10]:

$$\hat{H} = \hat{H}_{\text{BIC}} + \hat{H}_{\text{Ph}} + \hat{H}_{\text{int}} + \hat{H}_{\text{dia}} \quad (2)$$

$\hat{H}_{\text{BIC}}$ and $\hat{H}_{\text{Ph}}$ are the bare photon and phonon Hamiltonians, respectively. $\hat{H}_{\text{int}}$ is the light-matter interaction term, where the coupling strength is represented by $g$. $\hat{H}_{\text{dia}}$ is the $A^2$ term, which is the quadratic term of the vector potential A of the light field. Diagonalizing this Hamiltonian via a 4×4 matrix **M**, where $\mathbf{M}\vec{V} = \omega\vec{V}$ and $\det|\mathbf{M} - \omega\mathbf{I}| = 0$, yields the polariton eigenfrequencies $\omega_\pm$ (see Note 3, Supplementary Information). The resonance frequency of BIC and phonon are $\omega_{\text{BIC}}$ and $\omega_{\text{Ph}}$, and damping rate of BIC and phonon are $\gamma_{\text{BIC}} = 0.18$ THz and $\gamma_{Ph} = 0.20$ THz, the properties of the BICs and phonons are extracted from measurement (Figures 2b and c). And $g$ is the coupling strength between the two coupled elements. The corresponding Rabi splitting is defined as $\Omega_R$ by substituting

$\omega_+ - \omega_-$. As shown in Figure 2d, the interaction between the BICs and the phonons leads to a hybridization effect, resulting in the emergence of new resonant peaks that are characteristic of phonon-polaritons. The transmission amplitude shows clear anticrossing behaviour, which is an essential feature of strong coupling systems. It indicates that the BICs and phonons have strongly coupled to form new phonon-polariton modes. The transmission amplitude of the polariton modes shown in Figure 2d was analyzed according to Eq. 2. The minimum of the splitting corresponds precisely to the condition where $\omega_{\text{BIC}} = \omega_{\text{Ph}} = 0.95$ THz (Figure 2e). This result indicates that the system's Rabi splitting is 0.28 THz, which is correspond to 1.15 meV. According to established criteria of strong coupling [61] $c = 2g/\sqrt{(\gamma_{\text{BIC}}^2 + \gamma_{\text{Ph}}^2)/2} > 1$, which leads to $c = 1.47 > 1$. Obviously, the experimental results substantially exceed the required conditions, confirming that the coupling strength of our Au-MAPbI$_3$ hybrid metasurfaces indeed reach the criteria of strong coupling. Furthermore, the experimental results align well with the full Hamiltonian results (Figure S6, Supplementary Information), providing additional validation for our findings. The formation of these phonon-polariton modes enhances the light-matter interaction, leading to increased field localization and modified dispersion characteristics.

## 3. Tailoring the ultrastrong coupling in Au-MAPbI$_3$ hybrid metasurfaces

The general definition of Rabi splitting [16] is

$$\hbar\Omega_{\text{R}} = \hbar\Omega\sqrt{N} \propto \sqrt{\frac{N}{V_{\text{eff}}}} \qquad (3)$$

where $N$ is the number of oscillators participating in the coupling, and $\hbar\Omega$ represents the contribution from each oscillator. In strong coupling systems, the coupling strength is increased as the mode volume of BIC (see Note 4, Supplementary Information) decreases according to Eq. 3. In situations where it is challenging to increase the maximum electric field strength, compressing the mode volume by reducing the period of the metasurface can effectively enhance the Rabi splitting. So, the metasurfaces are designed with different periods by varying the parameter $d = P_x - L_x$ to tune the coupling strength by changing the mode volume. A series of different parameters $d$ from 50 to 2 μm are selected based on the metasurface with $d = 16$ μm in Figure 2d. $P_y - L_y$ is scaled proportionally with $d$, while other geometrical parameters of the metasurface remain constant according to the structural parameters from Figure 1b. In this case, the resonance of the BIC modes can match with phonons, which are in the optimal position for generating phonon-polaritons.

The transmission amplitude of the samples is shown in Figure 3a. When the period of the metasurface increases as $d$ increases from 16 to 50 μm, the Rabi splitting is modulated from 0.28 THz to 0.26 THz, indicating a slight change. When $d$ is less than 16 μm, the Rabi splitting is significantly enhanced as $d$ decreases. Specifically, as $d$ is reduced from 16 to 2 μm, the Rabi splitting is modulated from 0.28 THz to 0.46 THz. As a result, the Rabi splitting of the phonon-polariton is increased as $d$ decreases, with a modulation range from 28% to 48.4% of the phonon frequency. By extracting the peak frequencies of two branches from each transmission spectrum, the relationship between Rabi splitting and $d$ can be fitted. By simulation the electric near-fields in one unit cell with Lumerical, the mode volume ($V_{eff}$) was calculated as a function of the parameter $d$ (2 μm -16 μm). (see Note 4, Supplementary Information). Figure S7 quantitatively establishes the inverse proportionality between mode volume $V_{eff}$ and interlayer spacing $d$, showing a 62% reduction in $V_{eff}$ as $d$ decreases from 16 μm to 2 μm. This trend aligns with the $d$-dependent variation in Rabi splitting intensity shown in Figure 3b. In addition, given that $d$ is larger than the particle size observed after the crystallization of the MAPbI$_3$, the number of participating phonons remains at saturation levels, which means the influence of the number of phonons can be neglected. Besides, we demonstrate that BIC asymmetry exerts minimal influence on Rabi splitting values in phonon-polariton systems due to intrinsic-loss-dominated electric field enhancement limitations and high-order BIC mode in metallic metasurfaces, despite the strong coupling criterion parameter c increasing from 1.40 to 2.24 (see Note 5, Supplementary Information). Therefore, it is sufficient only to focus on the relationship between the Rabi splitting and mode volume $V_{\text{eff}}$.

To obtain the tunability of the phonon-polariton in the metasurface with the maximum Rabi splitting, 5 samples are prepared based on the metasurfaces with $d = 2$ $\mu$m. The resonance of BIC modes is tuned from 0.7 to 1.2 THz by changing the scaling factor $S$. Figure 3c presents the pseudo-color map derived from simulations incorporating both the BIC configuration and dual-phonon perovskite oscillator parameters (0.95+1.85 THz), with dark markers indicating experimentally extracted peak positions of the upper and lower energy branches. Clearly, anticrossing curves (color map in Figure 3c) were obtained by simulation, which means the frequency of phonon-polariton is tunable. The minimum frequency difference between the two polariton branches is 0.46 THz, occurring at the position where $\omega_{\text{BIC}} = \omega_{\text{Ph}}$. Consequently, the Rabi splitting produced by the sample is 48.4% of $\omega_{\text{Ph}}$, placing the system well within the regime of ultrastrong coupling. This value represents one of the highest coupling strengths achieved in terahertz phonon-polariton systems to date.

While excellent agreement exists between experimental measurements and numerical simulations, the observed position of the upper energy branch deviates from theoretical predictions based on the single-phonon (0.95 THz) Hamiltonian. This inconsistency suggests the presence of additional coupling mechanisms beyond the initial model. Detailed analysis in the Note 6 (Supplementary Information) indicates significant perturbation from the 1.85 THz phonon mode, which exhibits the spectral overlap with the coupled system[62]. Theory and simulations incorporating both phonons yield Rabi splitting values closer to experimental observations, confirming the necessity of concerning 1.85 THz phonon.

**4. Wavelet analysis process**

The wavelet transform is a powerful time-frequency analysis tool that overcomes the inherent trade-off between temporal and spectral resolution in conventional Fourier transforms. Unlike Fourier analysis, which decomposes a signal into infinite sinusoidal waves, wavelet transform uses localized basis functions ("wavelets") scaled and shifted across the time domain. This enables precise resolution of transient features in non-stationary signals like THz pulses. We employed the generalized Morse wavelet [63]. The generalized Morse wavelet in the Fourier domain is [64]

$$\psi_{\beta,\gamma}(\omega) = U(\omega) a_{\beta,\gamma} \omega^{\beta} e^{-\omega^{\gamma}} \qquad (4)$$

where $U(\omega)$ is the unit step function, $a_{\beta,\gamma}$ is a normalizing constant, $\beta$ controls the time-domain decay rate (related to compactness), and $\gamma$ characterizes the symmetry of the Morse wavelet.

Wavelet analysis yields the instantaneous spectrum for each delay in the mode evolution time trace, providing balanced results in both time and frequency domains. This method can be introduced as a powerful tool for analyzing the temporal information of THz emission and detection.

**5. Time evolution of electric mode via wavelet analysis**

To analyze the modes evolution of the generated phonon-polariton in this strong coupling system, THz fields passing through metasurfaces with representative periods are selected for the wavelet transform. As depicted in Figure 4a, the instantaneous spectrums of metasurfaces with four different parameters $d$ are generated by wavelet analysis (details see Figure S13, Supplementary Information), which is performed with the generalized Morse wavelet [63] with $\beta = 60$, $\gamma = 3$ (see Note 7, Supplementary Information). The zero delay corresponds to the peak electric field of the terahertz wave prior to coupling with the material. The depth information in figure 4a represents the absolute amplitude of the normalized THz field, showing the frequency distribution at different times as the THz wave passes through the metasurface. The amplitude of the THz field attenuates to zero in 10 ps, indicating the rapid decay time for the polariton modes. This observation aligns with the classical expression for mode evolution $Q = 2\pi t/2T = \omega_0 t/2$, where $t$ represents the oscillator lifetime, consistent with the damping rate derived from the results in Figure 3a. The modes in the directly obtained time-frequency spectrum are not clear enough, due to the influence of THz waves that do not participate in coupling. Therefore, in Figure 4b, clear modes evolution spectrum of the phonon-polariton are obtained by subtracting the backgrounds of Figure 4a. The three highlighted areas in Figure 4b, from top to bottom represent the upper polariton branch, the phonon mode, and the lower polariton branch respectively. The highlighted mode frequencies in Figure 4b correspond directly to the positions of the valley in the transmission amplitude spectrum in Figure 3a.

Notably, Figure 4a contain time evolution information of modes that cannot be obtained from the frequency domain spectrum through the Fast Fourier Transform (FFT). From the comparison of Figure 4b, it can be observed that the phonon mode appears 2 ps earlier as $d$ decreases. To further analyze the reasons for the temporal variations of the phonon modes in different samples, as shown in Figures 4c,d, the intensity and phase evolution information at $\omega = 0.95$ THz are obtained by processing the time-frequency spectrums for different parameters $d$. It is observed notable changes in the phase of THz field in Figure 4d. Specifically, both the THz field intensity and phase of phonon mode exhibit abrupt transitions, with the timing of these transitions occurring earlier as $d$ decreases. Furthermore, the wavepacket after the transition is primarily attributed to phonon re-emission (see Note 8, Supplementary Information). By comparing the THz field intensity of the re-emitted phonon portion in Figure 4c, it can be observed that the phonon radiation from the hybrid metasurface with parameter $d = 2\ \mu m$ is 10 dB higher than that with $d = 50\ \mu m$. It is concluded that as $d$ decreases, the coupling strength increases, and the radiative intensity of the phonon increases by 10 dB. This highlights the essence of enhanced coupling strength, which involves maximizing the proportion of radiative losses in the presence of material dissipation while minimizing the proportion of absorption losses. Similarly, the phase of the upper and lower polariton branches is significantly changed evolved over time, which reveals the process of energy absorption and re-emission within strong coupling systems (Figure S15, Supplementary Information).

To verify the role of photon-phonon hybridization, we conducted experiments at finite detuning ($\Delta = 0, 0.1, 0.2$ THz) while maintaining $d = 2\ \mu m$. As shown in Figure 5, detuning significantly reduced the delayed phonon re-emission intensity (by ~5 dB). In contrast, the hybrid system at zero detuning exhibited 20 dB stronger emission than bare perovskite films, confirming enhanced radiative efficiency via polariton. At zero detuning, complete hybridization channels energy exchange between the BIC mode and phonon, maximizing radiative losses. Conversely, detuning disrupts resonance, localizing energy in lossy phonon modes and increasing absorption losses. This behavior underscores the coherent coupling mechanism inherent in the phonon-polariton system, where the interactions between the phonons and the polaritons lead to the formation of these two modes. This behavior underscores the coherent coupling mechanism inherent in the phonon-polariton system, where the interactions between the phonons and the polaritons lead to the formation of these two modes.

**Discussion**

In summary, we demonstrate the ultrastrong coupling in Au-MAPbI$_3$ hybrid metasurfaces based on metallic BICs. The Rabi splitting can be continuously adjusted from 28% to 48.4% of the phonon frequency by modifying the period of the Au metasurfaces, thus reaching the ultrastrong coupling regime. Through wavelet transform, we successfully extracted the mode evolution information of the terahertz phonon-polariton. It indicates that the phonon radiation intensity of the MAPbI$_3$ film is enhanced with increasing coupling strength. Our findings provide critical insights into the interaction between the metallic BICs and the phonons of perovskite. The methodologies and insights gained can be applied to explore multi-mode coupling phenomena across various platforms and materials. In addition, Perovskite materials exhibit intrinsic semiconductor properties, making them promising candidates for optoelectronic applications. This perovskite-based platform demonstrates significant potential for further optoelectronic modulation in ultrastrong coupling regime [20,65], with promising applications in topological engineering [30], Ultrafast modulator [66] and Electronic transport [67]. This research fosters the development of innovative photonic devices, and opens new pathways for future investigations into polaritons and strong coupling effects in the terahertz regime.

**Materials and Methods**

**Sample fabrication**

The Au metasurfaces, with a thickness of 200 nm, were fabricated on a quartz substrate using UV lithography and Electron Beam Evaporation (EBE). The prepared samples were subsequently subjected to treatment in an ultraviolet ozone cleaner to thoroughly eliminate surface organic contaminants. Perovskite thin films were created using the solution spin coating method (Figure S18, Supplementary Information). A mixture of MAI and PbI$_2$ was dissolved in a solution of N, N-dimethylformamide (DMF) and dimethyl sulfoxide (DMSO) in a 1:1 ratio. The solution was stirred on a magnetic stirrer at 60°C for 12 hours until fully dissolved. The solution was then filtered to obtain the MAPbI3 perovskite precursor solution. Then, 40 μL of the MAPbI3 precursor solution was spin-coat onto the metasurface at 1000 rpm for the first 10 seconds, followed by 6000 rpm for the next 30 seconds. 100 μL of chlorobenzene was dispensed onto the surface quickly and evenly. After spin coating, the sample was placed on a hot plate and annealed at 100°C for 15 minutes to complete the perovskite thin film preparation about 250 nm (see Figure S19, Supplementary Information). Lead iodide (PbI2) was purchased from the Tokyo Chemical Industry. Methylammonium iodide (MAI) was purchased from MaterWin company. N, N-Dimethylformamide (DMF), dimethyl sulfoxide (DMSO), and chlorobenzene (CB) were obtained from Aladdin.

**Experiment setup**

We conducted terahertz spectroscopy measurements using our independently developed terahertz time-domain spectroscopy (THz-TDS) system (Figure S20, Supplementary Information), and obtained detailed information about the terahertz field through Fast Fourier Transform (FFT) and wavelet analysis. The THz-TDS system is driven by a Ti-sapphire femtosecond laser, which generates laser pulses with a duration of 35 fs, a repetition rate of 1 kHz, and a center wavelength of 800 nm. The femtosecond laser beam is split into two paths, used for terahertz generation and detection respectively. We utilize efficient metallic spintronic emitters of ultra-broadband terahertz radiation to generate terahertz pulses based on the principle of ultrafast photoinduced spin currents [68]. The emitted terahertz pulse is polarized in the y-direction in Figure 1b, and passes through the sample before simultaneously impinges on a ZnTe crystal with another femtosecond laser pulse, where the detection of the terahertz pulse is accomplished via the electro-optic effect. To eliminate the impact of water molecules in the air on the experiment, the portion of the test system through which terahertz waves propagate is filled with dry air maintained at a humidity level below 5%.

**Acknowledgements**

This work was supported by National Key Research and Development Program of China (2022YFA1604403); National Natural Science Foundation of China (12274157, 12021004, 12274334, 11904271); Natural Science Foundation of Hubei Province (2023AFA076). Special thanks are given to the Analytical and Testing Center of HUST, the Center of Micro−Fabrication and Characterization (CMFC) of WNLO and LBTEK for the use of their facilities.

**Data availability**

The main data supporting the findings of this study are available within the article and its Supplementary Information files. Extra data are available from the corresponding author upon reasonable request.

**Competing interests**

The authors declare no conflicts of interest.

**Author contributions**

K. W., P. X. L. conceived the project. K.W., P. X. L. supervised the project. J.X.Y., X. B. H. and F.R.L. designed the experiments. J.X.Y., A.Y.F. and C.Z. performed the experiments. J.X.Y., L.Y.Z. and X. B. H. performed theoretical calculations and numerical simulations. J.X.Y., L.Y.Z., K.W., A.Y.F., C.Z., Z.J.H., Z.D.L., X.B.H. analyzed data. All authors discussed the results. J.X.Y., C.Z., and L.Y.Z. drafted the paper with the inputs from all authors.

## Captions

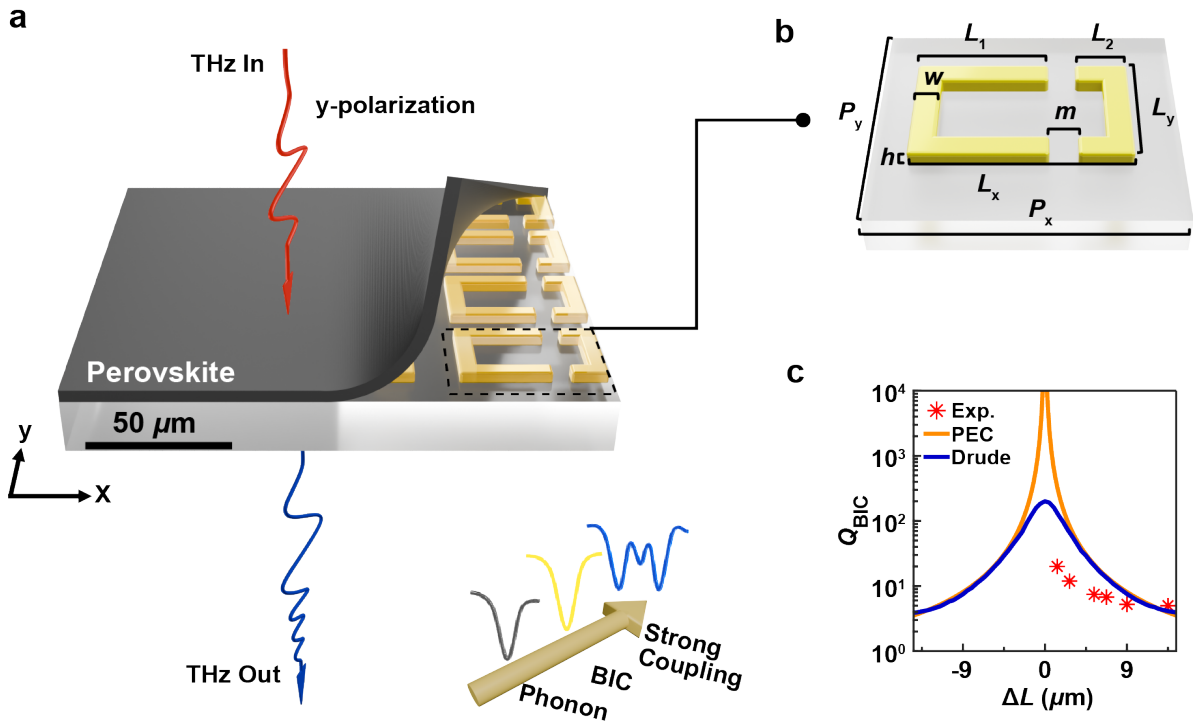

**Figure 1 | a,** Illustration of Au metasurfaces covered with a MAPbI$_3$ thin film. **b,** The structure of the BIC unit. The geometrical unit cell parameters are: $P_x$ = 66 μm, $P_y$ = 32 μm, $L_x$ = 50 μm, $L_y$ = 25 μm, $L_1$ = 39.5 μm, $m$ = 5 μm, $w$ = 5 μm, $h$ = 200 nm, with the period $d = P_x - L_x$. The tuning of the resonance position of the BIC modes is realized by introducing multiplicative scaling factor $S$, which scales the geometrical parameters of metasurfaces by multiplying all parameters by $S$. Terahertz waves pass through the sample in y-polarization and emerge from the quartz substrate side. (bottom right) Schematic diagram of strong coupling between phonons and BICs. **c,** Quality factor ($Q$) of the BIC modes varies with the degree of asymmetry. Red stars represent experimental data, while blue crosses represent the Drude model results, and simulations using parameters employed in the experiment. The orange solid line represents the ideal PEC case. The degree of asymmetry is defined by the expression $\Delta L = L_1 - L_2$.

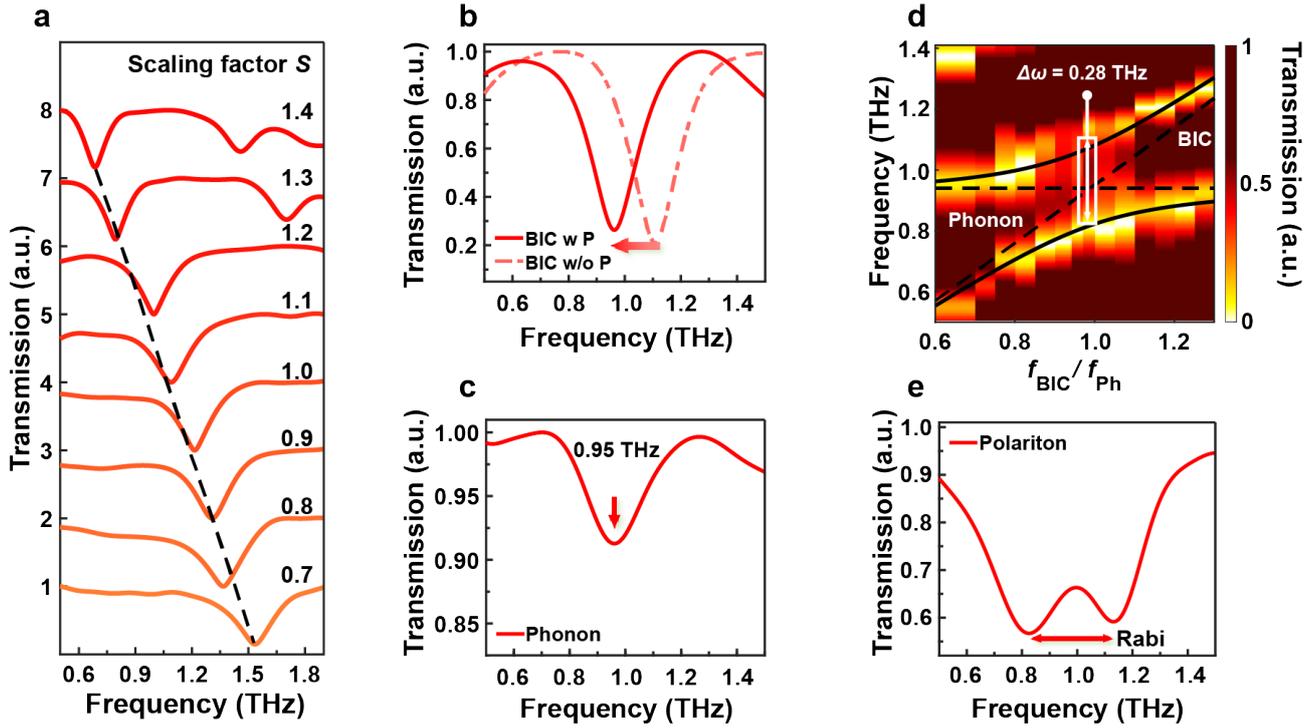

**Figure 2 | a,** Measured transmission amplitudes of BIC resonances across the frequency of perovskite phonon tuned via scaling the in-plane geometric parameters. The new parameters were derived by multiplying $S$ with the in-plane geometric parameters. $P_{xnew} = SP_x$, $P_{ynew} = SP_y$, $L_{xnew} = SL_x$, $L_{ynew} = SL_y$, $L_{1new} = SL_1$, $m_{new} = Sm$, $w_{new} = Sw$. **b,** Transmission amplitude of the Au metasurface coated with an amorphous perovskite film is represented by red solid line, while that of pare Au metasurface without perovskite is represented by a red dished line. **c,** The transmission amplitude of a MAPbI$_3$ film spin-coated on a quartz substrate. The arrow indicates the frequency of phonon resonance. **d,** Transmission amplitude of Au-MAPbI$_3$ metasurfaces for different scaling factors shows a characteristic and anticrossing mode pattern close to the MAPbI$_3$ phonon. **e,** Transmission amplitude of Au metasurface coated with a crystallized perovskite film while the resonance position of BICs is tuned to 0.95 THz, which reveals a Rabi splitting of 0.28 THz.

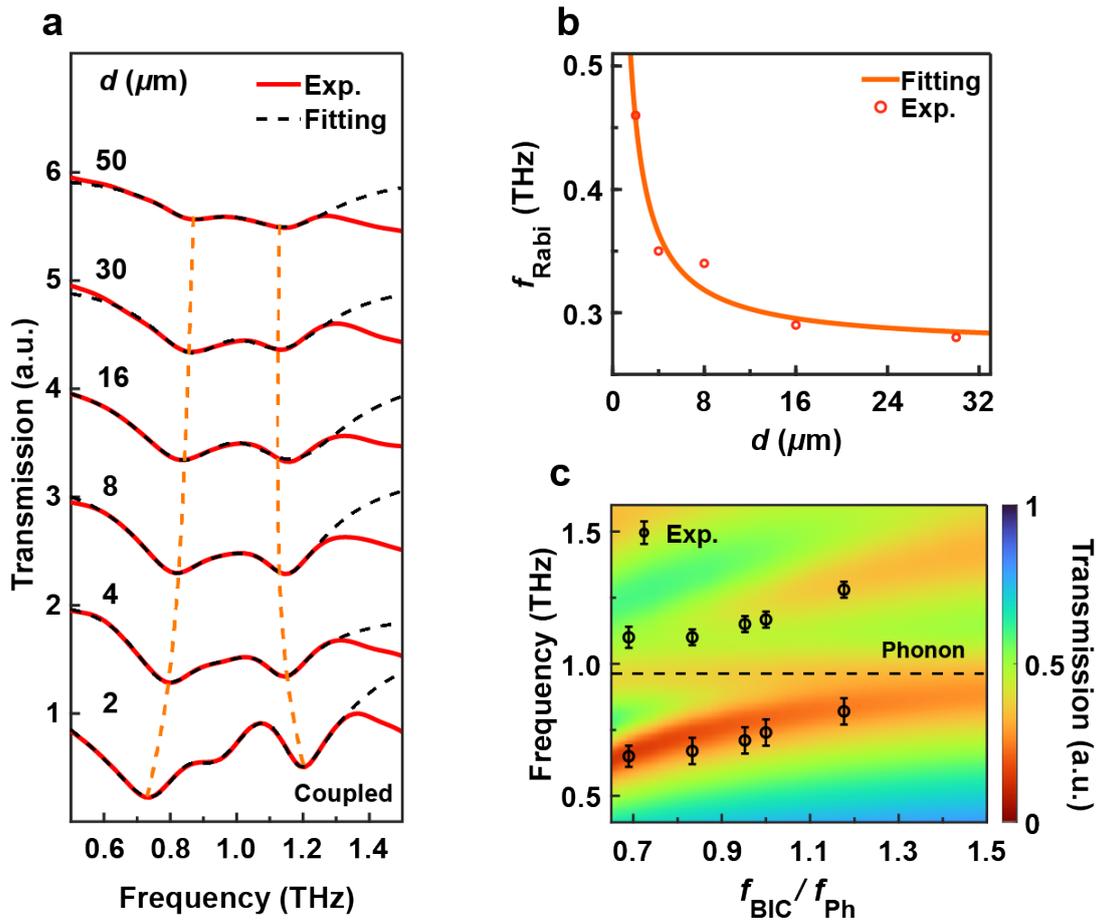

**Figure 3 | a,** Transmission amplitude that corresponds to the on-resonance case with the period $d$ changed, where $L_x = 50$ $\mu$m is constant. The red curve represents the experimental data, while the dark dashed line corresponds to the fit obtained from the experimental data. The peaks of polariton are connected by orange dashed lines. **b,** Experimentally measured correlation between Rabi splitting and parameter $d$, showing monotonic enhancement of splitting magnitude with decreasing $d$. **c,** The transmission spectra evolution with BIC cavity resonance frequency. The color map represents simulation results implementing both BIC configuration and dual-phonon perovskite oscillator parameters (0.95+1.85 THz), while dark markers denote experimentally extracted peak positions of the upper and lower polariton branches.

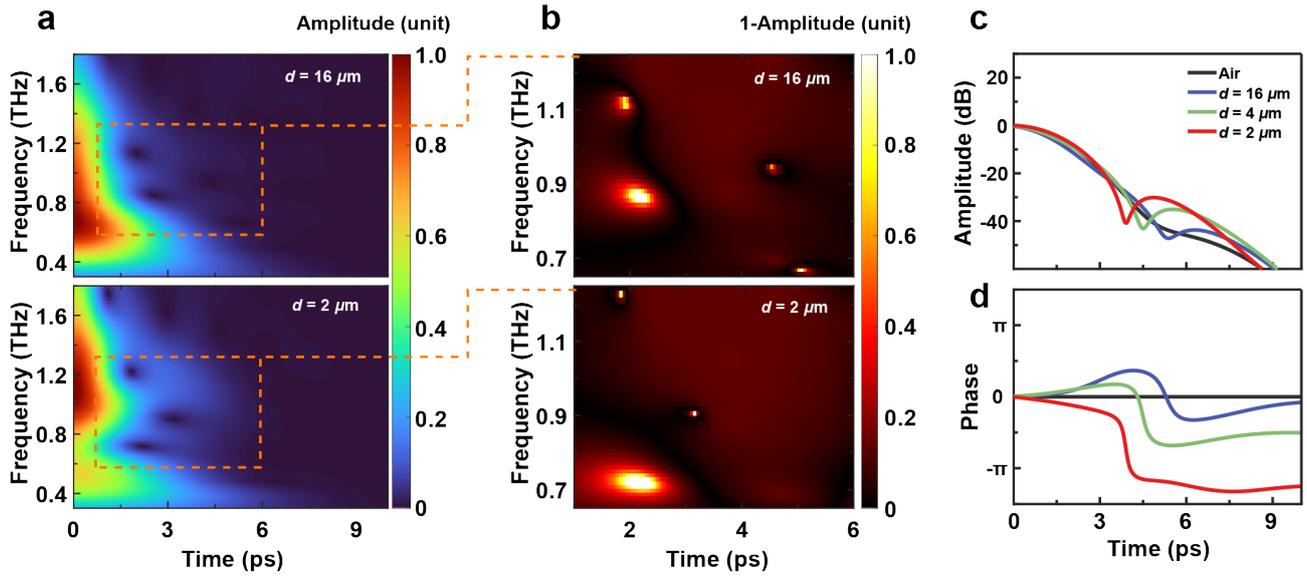

**Figure 4 | a ,** Time-evolved spectrum was obtained through the wavelet transform method with $d$ = 16 μm, $d$ = 2 μm. **b** Time-frequency domain maps of the regions $d$ = 16 μm and 2 μm after background removal within the orange dashed boxes. **c-d,** Time evolution of **(c)** intensity and **(d)** phase of the terahertz field at $\omega_{Ph}$ = 0.95 THz for different $d$ parameters.

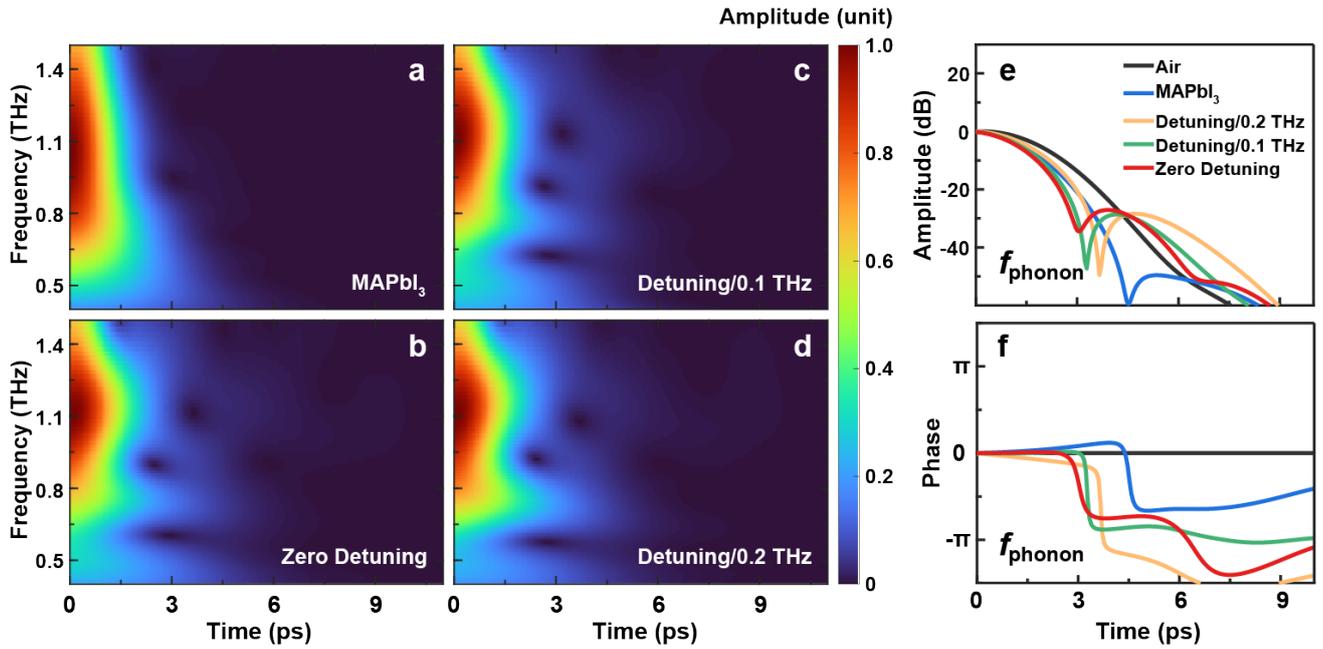

**Figure 5 | a-d,** Wavelet transform analysis of time-domain signals reconstructed from incident terahertz wavepackets and three coupled decaying oscillators, corresponding to the upper polariton branch, lower polariton branch, and phonon mode. Time-frequency maps are normalized to the incident pulse amplitude. **(a)** bare MAPbI$_3$ thin film. **(b)** zero detuning. **(c)** finite detuning ($\varDelta =$ 0.1 THz). **(d)** finite detuning ($\varDelta =$ 0.2 THz). **e-f,** Time evolution of **(e)** intensity and **(f)** phase of the terahertz field at $\omega_{Ph} = 0.95$ THz for air, bare MAPbI$_3$ and hybrid metasurface at the finite detuning with the same $d$ parameter.